\documentclass[superscriptaddress,aps,pra,nofootinbib,onecolumn,twoside,reprint,letterpaper,showpacs,10pt,floatfix]{revtex4-2}

\usepackage{graphicx}
\usepackage{dcolumn}
\usepackage{bm}
\usepackage{amsmath}
\usepackage{xcolor}
\usepackage{hyperref}
\usepackage{tablefootnote}

\usepackage{caption}
\usepackage{subfig}

\newcommand{\fund}{TM$_{010}$}
\newcommand{\ffund}{f_{010}}
\newcommand{\gagg}{g_{a\gamma\gamma}}
              
\begin{document}

\title{High-volume tunable resonator for axion searches above 7~GHz}

\author{Taj A. Dyson}
\affiliation{%
 Stanford University, Stanford, CA 94305, USA
}%
\author{Chelsea L. Bartram}
\affiliation{%
 SLAC National Accelerator Laboratory, 2575 Sand Hill Road,
Menlo Park, CA 94025, USA
}%
\author{Ashley Davidson}
\author{Jonah B. Ezekiel}
\author{Laura M. Futamura}
\author{Tongtian Liu}
\affiliation{%
 Stanford University, Stanford, CA 94305, USA
}%
\author{Chao-Lin Kuo}
\affiliation{%
 Stanford University, Stanford, CA 94305, USA
}%
\affiliation{%
 SLAC National Accelerator Laboratory, 2575 Sand Hill Road,
Menlo Park, CA 94025, USA
}%
\date{\today}

\begin{abstract}
\noindent
We present results from an experimental demonstration of a tunable thin-shell axion haloscope whose geometry decouples its overall volume from its resonant frequency, thereby evading the steep sensitivity degradation at high frequencies. An aluminum 2.6~L ($41\,\lambda^3$) prototype which tunes from 7.1 to 8.0~GHz 
was fabricated and characterized at room temperature. An axion-sensitive, straightforwardly tunable TM$_{010}$ mode is clearly identified with a room temperature quality factor, $Q$, of $\sim$5,000.
The on-resonance $E$-field distribution is mapped and found to agree with numerical calculations. Anticipating future cryogenic operation, we develop an alignment protocol relying only on rf measurements of the cavity, maintaining a form factor of 0.57 across the full tuning range.
These measurements demonstrate the feasibility of cavity-based haloscopes with operating volume $V\gg\lambda^3$. We discuss plans for future development and the parameters required for a thin-shell haloscope exploring the post-inflationary axion parameter space ($\sim$4 to $\sim$30~GHz) at DFSZ sensitivity.
\end{abstract}
\maketitle

{\em Introduction.}---A range of astrophysical and cosmological observations point to the existence of dark matter (DM), but the particles of which it consists have remained elusive \cite{1970ApJ...159..379R, Planck2020, Bertone_2005, Clowe_2006}. One of the most promising candidates for DM is the axion \cite{abbott83,preskill83,dine83}, which also solves the long-standing strong CP problem in particle physics \cite{PecciQuinn1977, Wilczek1978, Weinberg1978}. 

A well-established type of axion search experiment is the \emph{haloscope}, which consists of a resonating cavity immersed in a strong magnetic field \cite{sikivie1, ADMX:2021nhd, haystac21,capp}. The magnetic field interacts with putative ambient axions with coupling constant $\gagg$, converting them into a weak photon signal (e.g., \cite{marsh}). This signal resonantly drives the cavity at a frequency equal (in natural units) to the axion's mass, an unknown parameter. In this Letter, we report an experimental validation of a novel axion haloscope design that can extend axion sensitivity to frequencies above $\sim 4$~GHz. 


The axion parameter space is scanned by tuning the resonant frequency $f_0$ of the haloscope. Integrating for longer at a certain $f_0$ tightens the upper bound on the coupling constant $\gagg$ 
for DM of mass $m_\mathrm{DM} = h f_0/c^2$. The rate at which a haloscope can scan across $m_\mathrm{DM}$ while reaching a given upper bound on the coupling is known as its scan rate, $R$. Considering only properties of the haloscope's resonator,
\begin{equation}\label{eq:scan}
    R \propto V^2C^2Q,
\end{equation}
where $V$ is the resonator's volume, $Q$ is its quality factor, and $C$ is the form factor of the axion-sensitive resonating mode, defined as
\begin{equation}
\label{eq:C}
    C = \frac{\lvert \int \mathrm{d}V \bm{E} \cdot \bm{B} \rvert^2}{B^2V\int \mathrm{d}V \lvert \bm{E} \rvert^2} = \frac{\lvert \int \mathrm{d}V E_z\rvert^2}{V\int \mathrm{d}V \lvert \bm{E} \rvert^2},
\end{equation}
where $\bm{B}$ is the external magnetic field and $\bm{E}$ is the electric field of the mode. The second equality in Eq. \ref{eq:C} holds for cavities placed in a solenoid where $\bm{B}=B\bm{\hat{z}}$, so $\bm{E}\cdot\bm{B} = E_zB$ and $B$ is uniform. The mode most sensitive to axions is that with $\bm{E}$ aligned with $\bm{B}$ and with wavelength such that one half-cycle occupies the entire cavity. 
At frequencies around 1 GHz, haloscopes based on cylindrical cavities \cite{ADMX:2021nhd,capp} have reached sensitivity to Dine-Fischler-Srednicki-Zhitnitsky (DFSZ) models \cite{DINE1981199,Zhitnitsky:1980tq}.
However, the active volume of a proportionally scaled cylindrical haloscope decreases rapidly as $f_0$ increases. 
Since the scan rate $R$ scales as $V^{-2}$, it becomes prohibitive to cover the frequency range predicted for post-inflationary axion scenarios ($f_0\gtrsim4$\,GHz) \cite{Borsanyi2016, abbott83,Buschmann2022, Klaer_2017, Kawasaki2015} down to the DFSZ limits.

In this Letter, we describe the first characterization of a prototype cavity implementing the `thin-shell' geometry proposed in \cite{Kuo_2020} and \cite{Kuo_2021} that alleviates the steep scaling of volume with frequency. Importantly, we show that the efficiency loss due to mode crossing in such an overmoded cavity is modest over a large frequency range.  We present an overview of the thin-shell geometry and detail its characterization. Finally, we discuss expansions to the design and the prospects for this high-frequency haloscope design.

\begin{figure}
    \centering
    \includegraphics[width=0.4 \columnwidth]{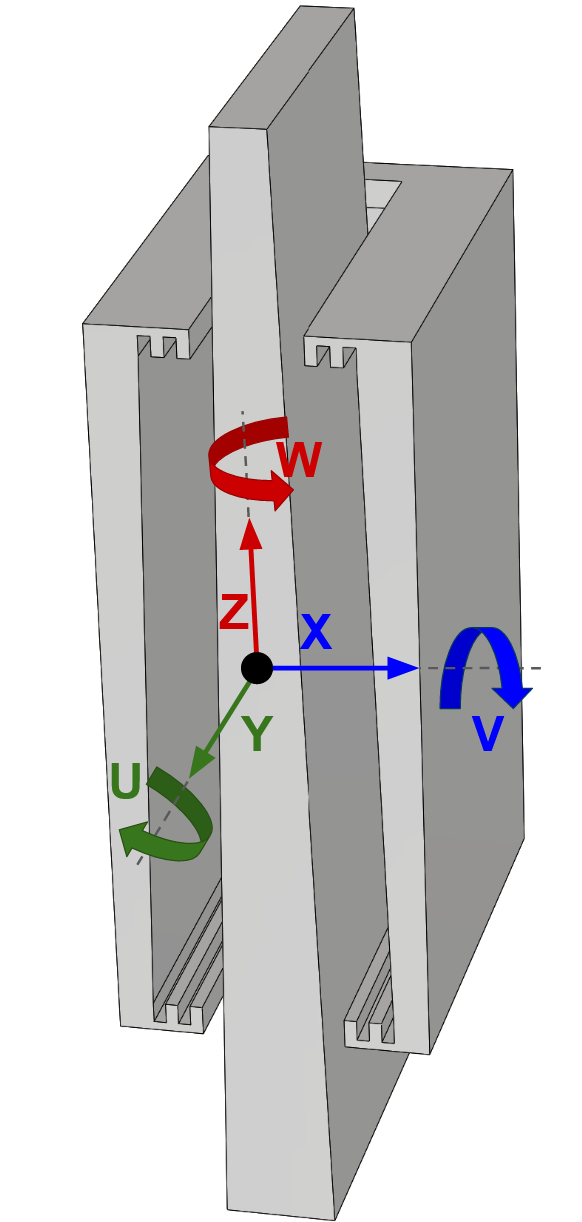}
    \caption{A cross-section through the center of the essential haloscope geometry, with the coordinate system used throughout this paper shown. The Z axis points vertically and the Y axis lies along the long edge of the wedge, with origin at the center of the wedge.}
    \label{fig:model}
\end{figure}

{\em The thin-shell geometry.}---It is possible to decouple the scan rate $R$ from $f_0$ by departing from the conventional cavity geometry. Reference \cite{Kuo_2021} provides a prescription to create {\em tunable} volume-filling thin-shell cavities with uniformly polarized eigenmodes. The simplest implementation of such a design consists of a rectangular wedge surrounded by a mechanically separate shell, shown in Fig. \ref{fig:model}, where the coordinates used throughout this paper are also defined. 

In the resulting cavity (a thin shell of vacuum space), standing waves in the X direction create a spatially uniform Z-polarized mode that maximizes $C$. In keeping with the terminology from cylindrical resonators, we call this the \fund\ mode. Its resonant frequency $\ffund$ is set by the gap width. The volume, on the other hand, can be scaled up by increasing the other dimensions to the size of the magnet bore without affecting $\ffund$. 

The cavity is tuned by moving the shell vertically along the wedge. Because the wedge and inner shell walls are tilted, such motion changes the width of the gap between them, thus tuning the \fund\ mode. Note that the wedge must be longer than the shell to allow tuning. The shell also has corrugated end caps, thus reducing radiative losses and improving $Q$ \cite{Kuo_2020}. As described in the reference, these corrugations are $\lambda/4$ deep and confine the fields to within the cavity. Note that for the \fund\ mode to be supported with appreciable $C$, the two parts of the resonator must be sufficiently well aligned.


{\em The prototype cavity.}---An aluminum thin-shell resonator was constructed and studied. The wedge measures 40~cm tall, and the shell is 24~cm tall by 24~cm wide and 7~cm deep. The shell consists of six flat plates and was assembled using precision parallels for alignment and spacing. Both the wedge and the inner surfaces of the shell were lapped down to a maximum deviation from the plane of 10~$\mu$m to improve $Q$. The surface flatness is verified using a capacitive displacement sensor from Micro-Epsilon. 

The rectangular wedge is mounted vertically on a Newport Hexapod (HXP100V6-MECA) six-axis positioner. The shell is mounted to a single-axis positioner colinear with the Z axis. Both positioners are mounted on a leveled, low-vibration optical table.  The resonator was designed to tune from 6.8 to 8.3~GHz.  However, the maximum travel of the Z-axis positioner is 10 cm, which limits the tuning range.

The rf response of the haloscope is probed with a 6.8~cm copper antenna mounted on an SMA feedthrough to the shell at the top on the $-$Y side of the cavity.
Cavity resonances are probed via reflection measurement of this antenna using a commercial vector network analyzer (VNA). The antenna is weakly coupled to the cavity ($\beta\approx0.1$), probing the unloaded $Q$ of the resonator. 

\begin{figure}
    \centering
    \includegraphics[width=\columnwidth, keepaspectratio]{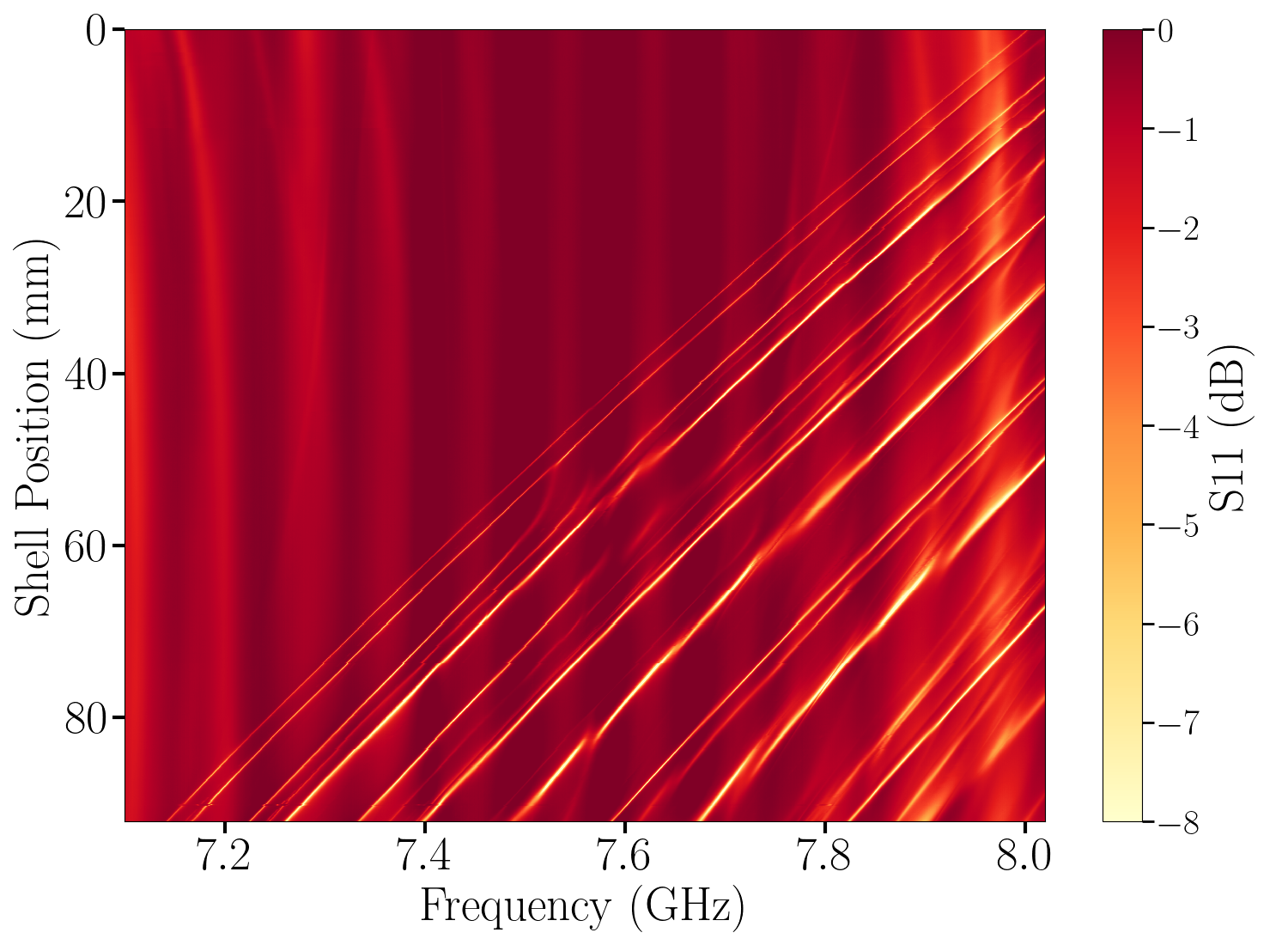}
    \caption{Mode map showing the tuning range of the prototype cavity explored in this study. Each row is an S11 sweep at a given shell Z position. The measured resonant modes are visible as bright lines (the vertical ripples are known VNA artifacts). The axion-sensitive \fund\ mode has the lowest frequency, and tunes uniformly from 7.1~GHz to 8~GHz. The wedge is aligned every 0.4~mm (3.9~MHz) using the procedure detailed in the main text.}
    \label{fig:zmap}
\end{figure}

\noindent A simple alignment using caliper measurements is sufficient to produce a resonance with the expected tuning properties. The observed frequency is 15.2~MHz lower than predicted (0.2\%), likely due to imperfect cavity construction.  We proceed to verify that the detected mode is indeed the axion-sensitive \fund\ mode against the background of spurious modes (see Fig.~\ref{fig:zmap}) by comparing its properties with those calculated using finite element analysis (FEA) by COMSOL-RF using two methods.

\begin{figure}
    \centering
    \includegraphics[width=\columnwidth]{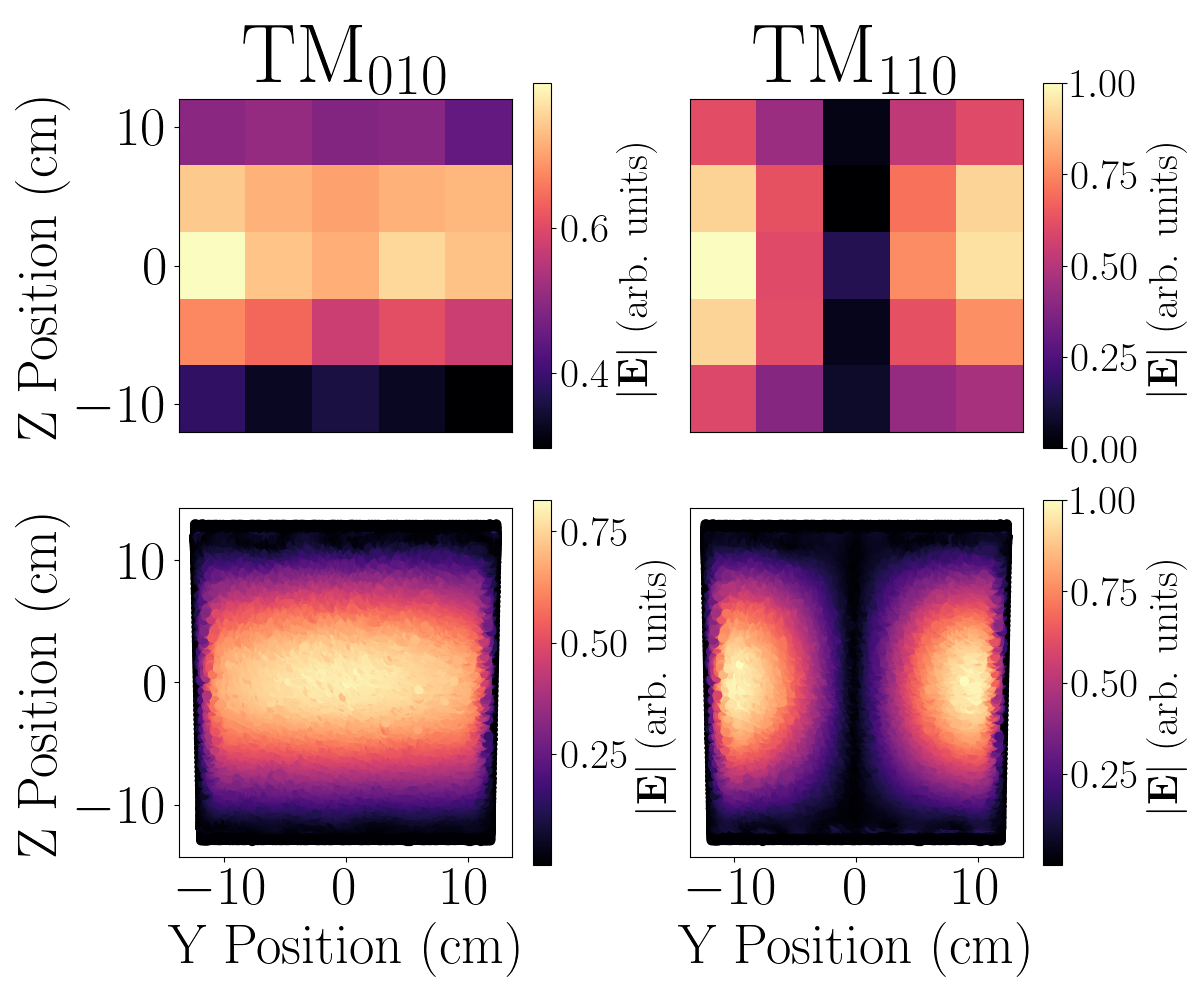}
    \caption{The left-hand column shows the \fund\ mode mapped with a dielectric disk (top) and using eigenfrequency FEA using COMSOL-RF (bottom). The right-hand column shows the same for the TM$_{110}$ mode. The measured and simulated plots use independent $|\bm{E}|$ scales. For details on the field mapping procedure, see the main text. The agreement between the measured and simulated field maps aids in identifying the \fund\ mode.}
    \label{fig:fieldmap}
\end{figure}

First, we measure the spatial distribution of the electric field inside the cavity using the ``bead perturbation'' technique \cite{bead52}. 
When a dielectric probe is placed in a resonating cavity, the frequency of each mode shifts proportionally to the square of the electric field of that mode in the probe's location.
In our experiment, the probes are Zotefoam HD30 disks, 19~mm in diameter and 14.3~mm thick. The choice of a larger probe with a low dielectric constant (over a more conventional probe consisting of a small bead with high $\epsilon$) is motivated by the large survey volume and the low spatial resolution required.  Maps of $\lvert \bm{E} \rvert$ within the cavity are produced by placing the probe at 25 evenly spaced points in each of the front ($+$X) and rear ($-$X) half of the cavity.
The first row of Fig. \ref{fig:fieldmap} shows the front field maps of the modes with the lowest and second lowest frequency, while the bottom row shows the matching field maps produced by FEA, that of the \fund\ and TM$_{110}$ modes.

\begin{figure}
    \centering
    \includegraphics[width=\columnwidth]{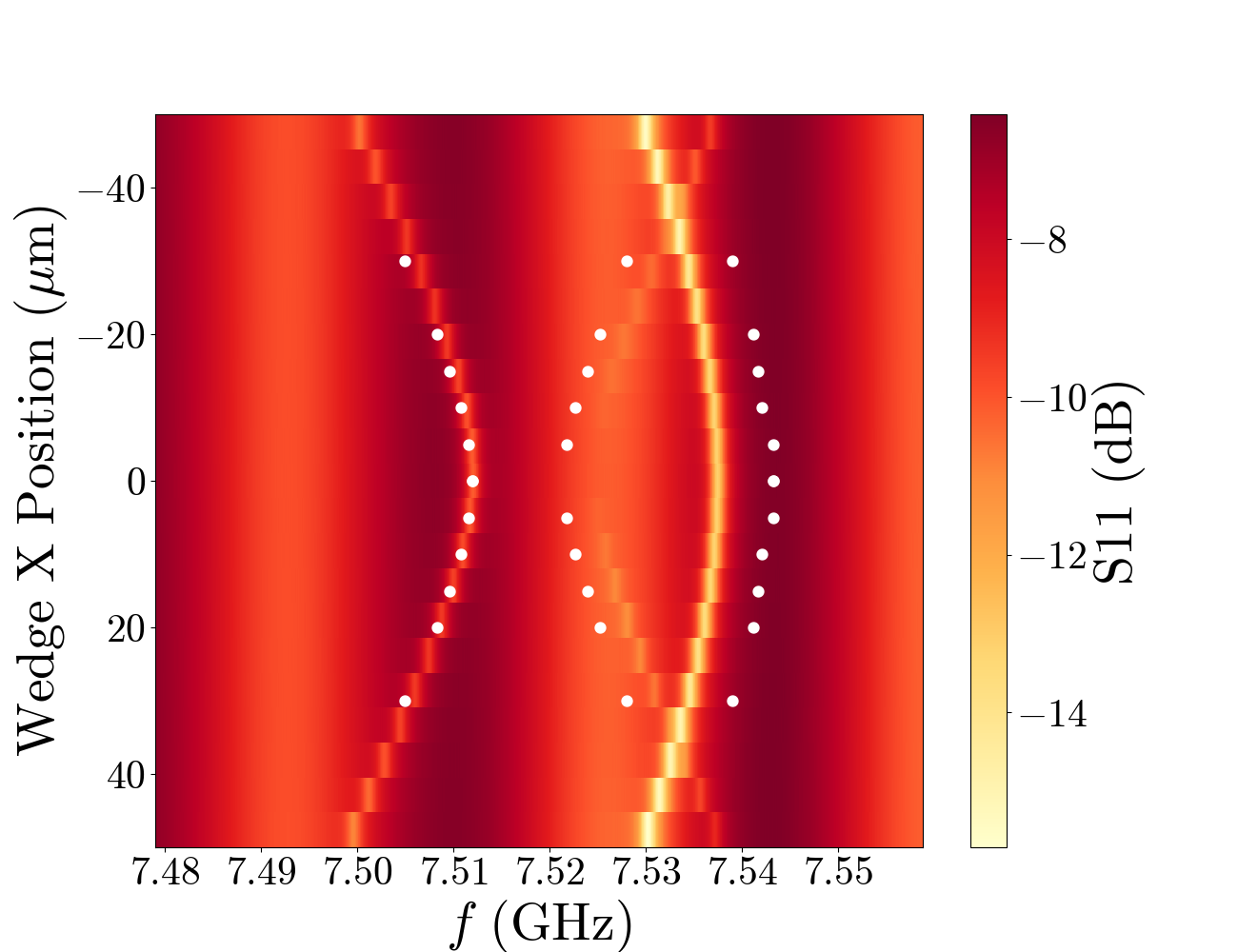}
    \caption{Mode map showing the effects of misaligning the wedge in the X direction. Each row is an S11 sweep at a given wedge position. The measured resonant modes are visible as curved bright lines (the vertical bars are known VNA calibration artifacts). The resonant frequencies given by FEA, shifted down by 15.23~MHz, are overplotted as white points.}
    \label{fig:modemap}
\end{figure}

Next, we corroborated these results by observing the responses of the modes' resonant frequencies to induced wedge misalignment. To do so, reflection measurements of the cavity are taken for several wedge positions along the X-axis, shown in Fig. \ref{fig:modemap}. In this figure, the resonant frequency given by FEA for the lowest three modes at several X positions is overplotted in white, uniformly shifted by 15.23~MHz to match the \fund\ mode's frequency to the reflection measurements. We observe excellent agreements between the measured and simulated responses to misalignment for the first three resonances. 

A feature of note in Fig.~\ref{fig:modemap} is the avoided crossing between the two lowest modes. This behavior is consistent with the expectation of two coupled harmonic oscillators -- the front ($+$X) and rear ($-$X) halves of the cavity. 
Furthermore, the next-lowest mode disappears when the wedge is aligned. This is expected because this mode corresponds to the anti-symmetric hybridization of the two halves. Since the antenna probe is located at the node, the coupling vanishes when the wedge is centered. 
We have thus associated the physical modes in the cavity with those simulated in FEA and confirmed that the axion-sensitive mode is supported in the cavity.

\begin{figure}
    \centering
    \includegraphics[width=\columnwidth]{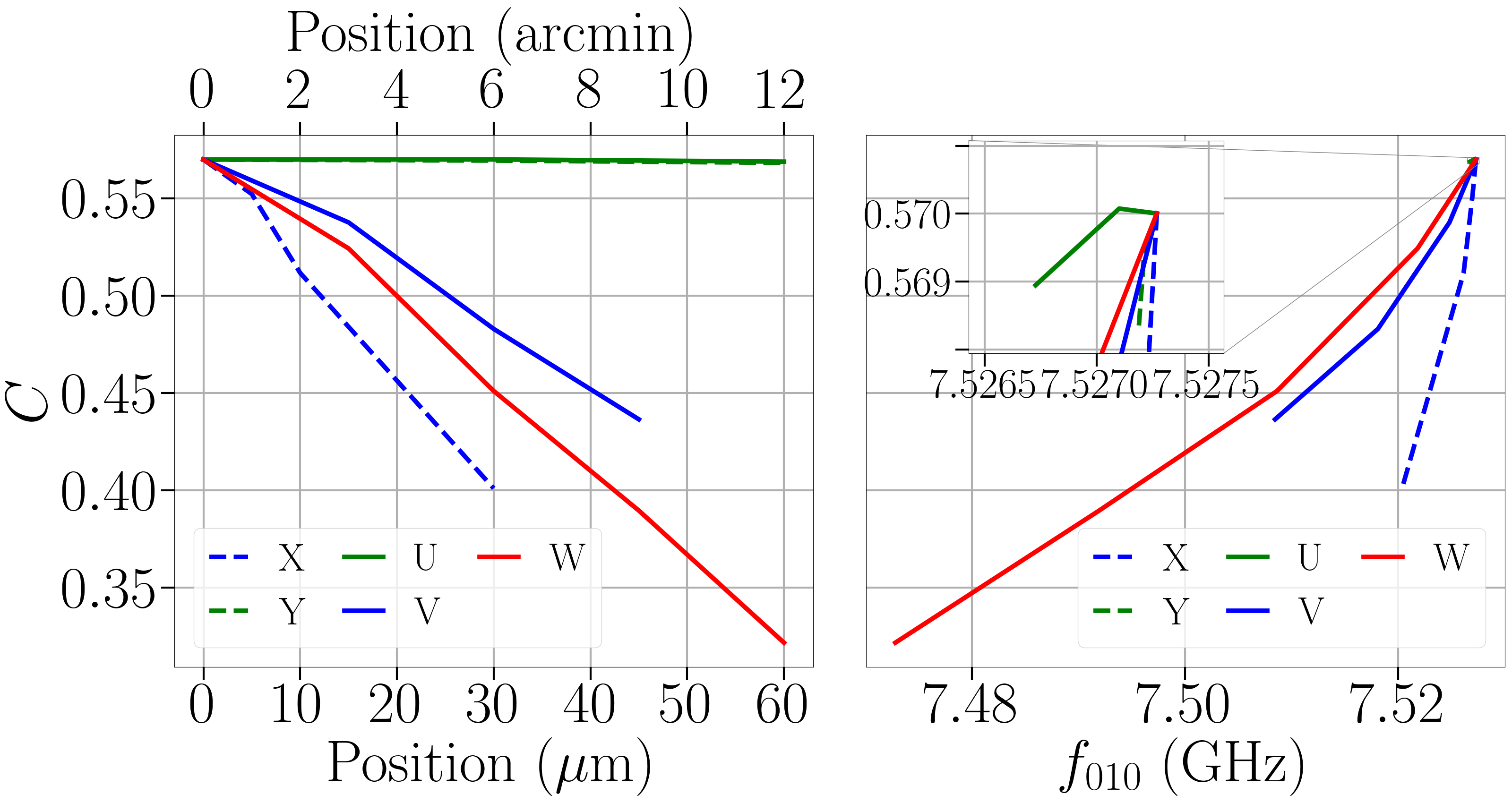}
    \caption{The left-hand panel shows $C$ as predicted by FEA as a function of wedge misalignment along each non-tuning axis. The right panel shows the effects of those same misalignments on $C$ as well as $\ffund$, showing that they decrease together as misalignment becomes more extreme in any direction. In the left-hand panel, dashed lines use the bottom position scale, solids use the top. Misalignment in the X, V, or W direction degrades $C$ significantly, while the effects of misalignment in the Y or U directions are negligible.}
    \label{fig:formfactor}
\end{figure}

Precise alignment of the wedge to the shell is important for optimal $C$ for the \fund\ mode. The simulated effects of wedge misalignment on $C$ are shown in the left-hand panel of Fig.~\ref{fig:formfactor}, while the right panel shows the effects of those same misalignments on $C$ and $\ffund$ together. When the wedge is misaligned, and $C$ is thus decreased, some part of the cavity must widen, driving $\ffund$ lower. This can also be seen in Fig. \ref{fig:modemap}: the aligned position has the highest $\ffund$. 
Since $\ffund$ decreases monotonically with $C$, aligning the wedge by maximizing $\ffund$ will also maximize $C$. This correspondence enables alignment without measurements beyond the usual rf characterization of the cavity, which is indispensable in cryogenic environments where other measurements are impractical. 

\begin{figure}
    \centering
    \includegraphics[width=0.8\columnwidth]{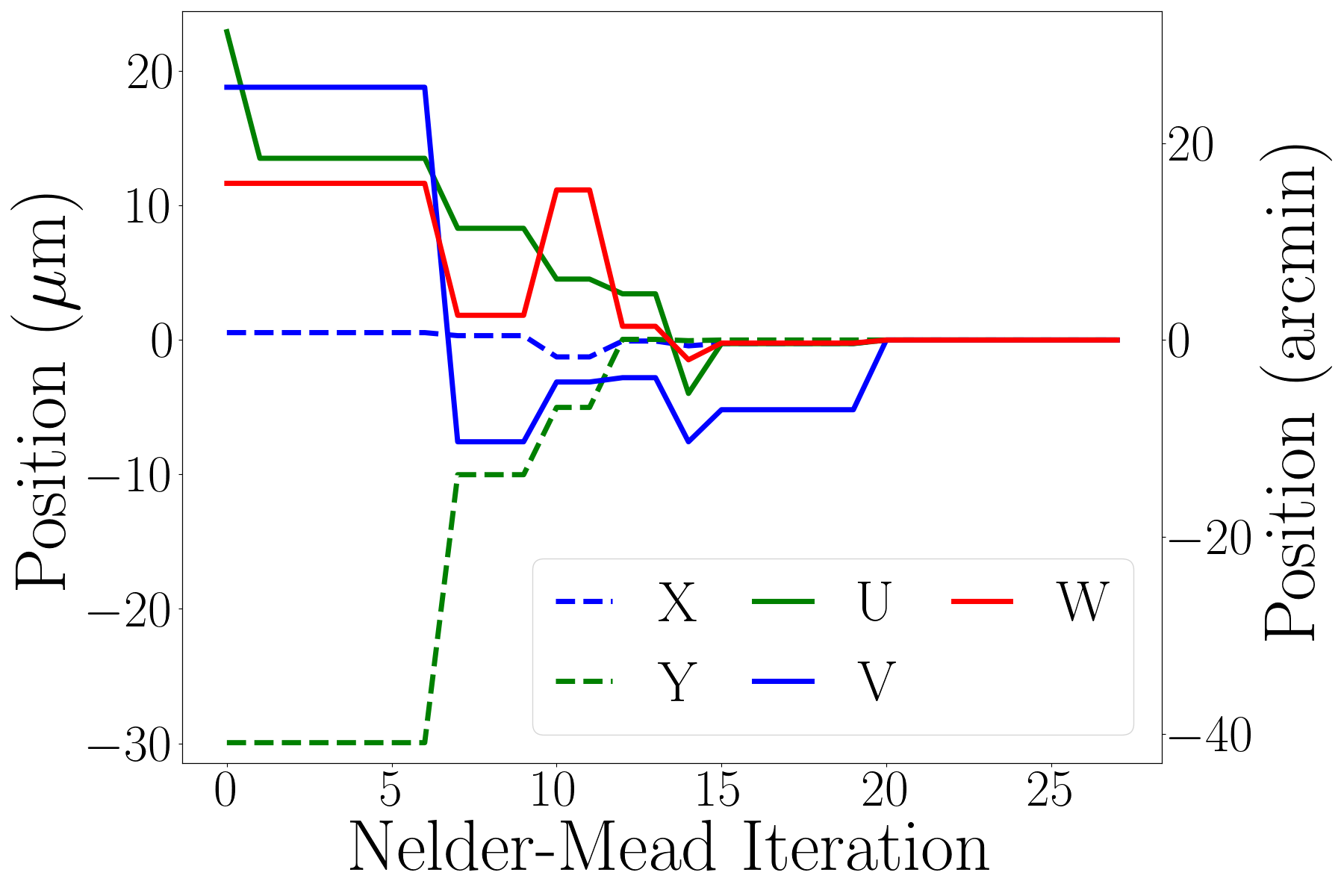}
    \caption{Position of the wedge as a function of NM iteration for a representative automatic alignment of the resonator. Dashed lines use the left axis, solid lines use the right axis.}
    \label{fig:NMalign}
\end{figure}

We developed an automated alignment algorithm to maximize $\ffund$ (and thus $C$) as a function of wedge position in each direction except Z (the tuning direction), including rotations. Given a position, the 6-axis hexapod places the wedge there. Then, an rf reflection measurement of the cavity with a VNA produces the cavity spectrum, from which $\ffund$ can be extracted. The above process is iterated, optimizing for maximum $\ffund$, until the optimization algorithm converges. A standard Nelder-Mead (NM) optimization algorithm was chosen to minimize the 5-dimensional function. 
The optimization is set to converge when all vertices of the parameter simplex are separated by no more than 1~$\mu$m for the linear dimensions and 3.6~arcseconds for the angular dimensions.  As can be seen in Fig. \ref{fig:modemap}, there are no false local maxima near the aligned position. The wedge's position during a representative alignment can be seen in Fig.~\ref{fig:NMalign}. Each function evaluation takes $\sim$1~s, and each alignment requires $\sim50$~evaluations ($\sim25$~NM iterations).  The successful implementation of the automated alignment is crucial in obtaining the data shown in Fig.~\ref{fig:zmap}, a key result of this Letter. 




The error on the aligned position of the wedge is estimated by aligning it 100 times, starting from uniformly random wedge positions. The error on the alignment position in each direction is then taken to be the standard deviation of the final aligned positions, and are recorded in Table~\ref{tab:errs}.

\begin{table}[]
\begin{tabular}{|c|c|c|c|c|c|}
\hline
\textbf{Axis}       & \begin{tabular}[c]{@{}c@{}}X\\ ($\mu$m)\end{tabular} & \begin{tabular}[c]{@{}c@{}}Y\\ ($\mu$m)\end{tabular} & \begin{tabular}[c]{@{}c@{}}U\\ (arcmin)\end{tabular} & \begin{tabular}[c]{@{}c@{}}V\\ (arcmin)\end{tabular} & \begin{tabular}[c]{@{}c@{}}W\\ (arcmin)\end{tabular} \\ \hline
$\bm{\sigma_x}$ & 2                                                    & 10                                                   & 0.4                                                    & 0.2                                                    & 0.1                                                     \\ \hline
\end{tabular}
\caption{Standard deviation of aligned position along each wedge axis from 100 NM alignments.}\label{tab:errs}
\end{table}

Since $C$ requires information about the direction of $\bm{E}$ inside the cavity, it cannot be measured directly. Instead, we calculate it from the FEA, which predicts that the fully aligned cavity has $C=0.57$. The positioning error of the wedge around the aligned position after alignment (Table \ref{tab:errs}) results in negligible degradation of $C$ (see Fig. \ref{fig:formfactor}, left-hand panel).


The cavity tunes linearly from 7.10 to 8.02~GHz, with no observed mode crossings (see Fig. \ref{fig:zmap}). Alignments were performed every 0.4~mm to keep $C>0.55$ throughout. Over this tuning range, four hours were spent on alignment, which is small compared to the expected data collection time during a future axion search. The demonstrated tuning range of $12\%$ is smaller than the design range of $20\%$ due to limitations in the mechanical setup and the coupling strength with the given antenna probe. 

At room temperature, the aluminum cavity's \fund\ mode has a median $Q$ of 4,700 throughout the full tuning range shown in Fig. \ref{fig:zmap}. 
FEA indicates that plating the cavity with copper and cooling to 100~mK will increase $Q$ to $20,000$, as radiative loss through the top and bottom gaps starts to become important.  

The cavity has volume $V=2.59$~L, or 41~$\lambda^3$, at 7.5~GHz. For a comparison with other experiments targeting a similar frequency range, see Table~\ref{tab:FOMs}.

\begin{table}[]
\begin{tabular}{|l|l|l|l|}
\hline
{\bf Experiment}      & $\bm{f_0}$ (GHz) & $\bm{V/\lambda^3}$ & $\bm{C^2V^2Q_L}$ (L$^2$) \\ \hline
This work (cryogenic) & 7.5              & 41                 & 15,000                   \\ \hline
HAYSTAC \cite{HAYSTAC:2023cam}              & 4.7              & 6.0                  & 2,600                    \\ \hline
CAPP quad-cell \cite{JEONG2018412}            & 5.89             & 8.2                & 3,100                     \\ \hline
HAYSTAC 7-rod \cite{Simanovskaia_2021}        & 6.5              & 17                 & 5,800                    \\ \hline
QUAX \cite{PhysRevApplied.17.054013}                 & 10.35            & 44                 & 1,200                    \\ \hline
GigaBREAD \cite{Knirck:2023jpu}            & 11.6              & 390              & 110                      \\ \hline
ORGAN \cite{PhysRevLett.132.031601}                & 26.5             & 11                 & 0.23                     \\ \hline
\end{tabular}
\caption{Comparison between this work and experiments targeting a similar frequency range.
In cases where only $Q_0$ is reported (this work, HAYSTAC 7-rod, and CAPP quad-cell), cavity coupling $\beta=2$ is assumed with $Q_L=Q_0/(\beta+1)$. GigaBREAD is a dish antenna, so equivalent $V=A\lambda/2$ where $A$ is its area, and for comparison $C$ and $Q_L$ are each taken to be $1$. The $Q_L$ quoted for this work is derived from the simulated $Q_0$ for cryogenic copper of 20,000.}
\label{tab:FOMs}
\end{table}

{\em Discussion.}---The axion scan rate $R$ can be further improved by copper-plating, cooling the haloscope to cryogenic temperatures, and stacking multiple wedge cavities (in the X direction) to form a larger volume. In the multi-wedge design, the sub-cavities are left open to each other, so a single axion-sensitive mode occupies the entire space. 

Efforts are being made to develop a meter-scale cavity for cryogenic operations to verify the expected improvements to $Q$ at low temperatures and tackle challenges relating to cryogenic positional control of the wedge. 
A three-wedge prototype is being developed to demonstrate the alignment of all wedges simultaneously. Additionally, an antenna scheme capable of coupling more strongly ($\beta\approx1$~to~$2$) to thin-shell cavities is under development. We baseline a science-grade 5--8 GHz haloscope on a copper-plated three-wedge design. 

\begin{figure}[h]
    \centering
    \includegraphics[width=\columnwidth]{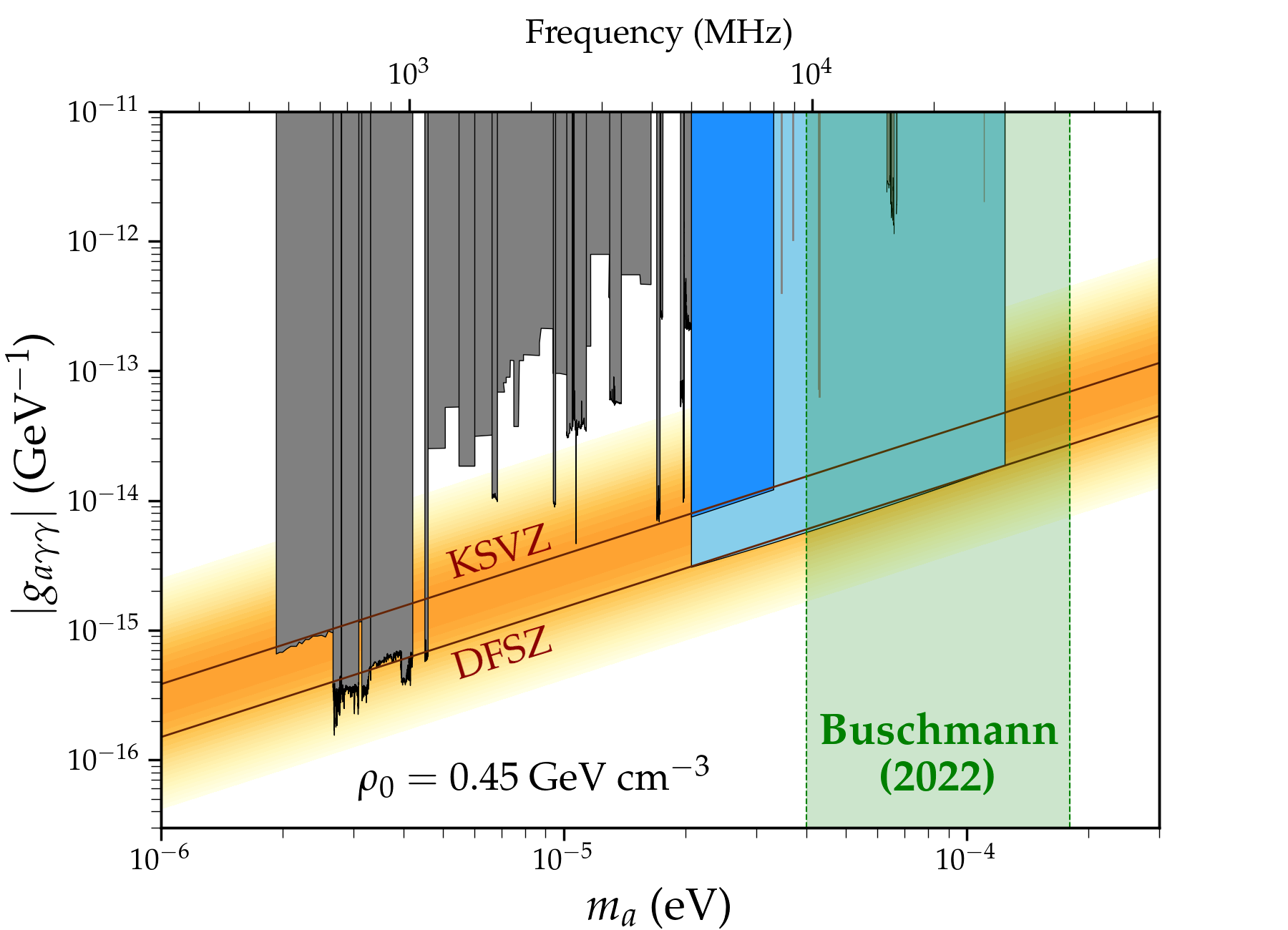}
    \caption{Discovery potential of thin-shell haloscopes. The green-shaded region is the post-inflationary axion mass range predicted by \cite{Buschmann2022}. A modest projection is in dark blue, reaching KSVZ sensitivity from 5-8~GHz, and a futuristic projection is in light blue, reaching DFSZ sensitivity from 5-30~GHz. Details of the parameters used are in the main text. The software used to produce this figure is adapted from \cite{ciaran_o_hare_2020_3932430}.}
    \label{fig:forecast}
\end{figure}

The discovery potential of such a cavity is shown in Fig. \ref{fig:forecast}, along with existing haloscopes' exclusions and the Kim-Shifman-Vainshtein-Zakharov (KSVZ) \cite{kim,shifman} and DFSZ \cite{DINE1981199,Zhitnitsky:1980tq} limits. In the following description, the frequency-dependent values $Q$ and $V$ are quoted at a fiducial frequency $f_\mathrm{fid}=5$~GHz, and $Q$ scales as $Q=Q_\mathrm{fid}(f_\mathrm{fid}/f)^2$ while $V$ scales according to the geometry of the cavity. The dark blue region is the parameter space excluded by an optimally coupled ($\beta=2$ \cite{ALKENANY201711}) three-wedge cavity, scaled to fit in a 1~m solenoid with a 0.6~m inner bore ($V=48$~L) obtaining $C=0.57$ and $Q=20,000$, cooled to 100~mK with a 300~mK first-stage amplifier (such as a Josephson parametric amplifier), immersed in an 8~T magnetic field, with 2~months of data collection time. The light blue region is a futuristic haloscope making use of a larger 10~T solenoid that is 2~m in length with a 0.8~m inner bore, containing a $V=116$~L cavity achieving $Q=100,000$ (again with $C=0.57$ and cooled to 100~mK) and 3.5~years of data collection time, assuming we employ a readout device with added noise at 1/2 of the standard quantum limit in the bandwidth of interest, such as a squeezed kinetic inductance traveling wave parametric amplifier \cite{Shu_2021} or single microwave photon detector \cite{Saclay_SMPD}.

{\em Conclusion.}---In this paper, we characterize a prototype thin-shell resonator, as proposed in \cite{Kuo_2020,Kuo_2021}, and find that it supports an axion-sensitive mode with appreciable form factor $C$ and $Q$ while achieving a $V/\lambda^3$ many times that of other experiments targeting a similar frequency range (see Table~\ref{tab:FOMs}). Furthermore, the resonator is tunable from 7.1-8~GHz via single-axis motion. The resonator consists of a thin shell of vacuum space in between a central wedge and its surrounding shell (see Fig.~\ref{fig:model}), and its $C$ is maximized when those two pieces are aligned. An automated alignment procedure was developed using only rf measurements of the mode and requiring only a modest amount of time to complete.

The prototype characterization demonstrates that the thin-shell haloscope design can have an operating volume $V/\lambda^3\gg1$ while retaining many desirable features of the successful cavity haloscope programs: mechanically robust high-$Q$ resonators, geometry compatible with a solenoid magnet, and a simple antenna readout. The wide frequency range covered shows that any efficiency loss due to mode crossing in such an overmoded cavity will be modest. Since for a haloscope $R\propto V^2$, the thin-shell cavity is a very promising approach to cover the important post-inflationary axion mass region. Compared to other innovative proposals to cover this range \cite{madmax_status_20,alpha}, a straightforward extension to this successful prototype is a particularly low-risk approach that should be considered seriously. Forecasts on the discovery potential of thin-shell haloscopes, in both the near and long term, are shown in Fig.~\ref{fig:forecast}.


\section*{Acknowledgements}

This material is based upon work supported by the National Science Foundation under Grant No. 2209576 and a KIPAC Innovation Award. We acknowledge the support of the Natural Sciences and Engineering Research Council of Canada (NSERC), 521528828. Chelsea Bartram and simulations and data acquisition advising were supported by the Department of Energy, Laboratory Directed Research and Development program at SLAC National Accelerator Laboratory, under contract DE-AC02-76SF00515 and as part of the Panofsky Fellowship awarded to Chelsea Bartram.

\bibliography{references}

\end{document}